%% file: main.tex
\documentclass[letterpaper]{article} 
\usepackage[]{aaai24}  
\usepackage{times}  
\usepackage{helvet}  
\usepackage{courier}  
\usepackage[hyphens]{url}  
\usepackage{graphicx} 
\usepackage{natbib}  
\usepackage{caption} 
\usepackage{algorithm}
\usepackage{algorithmic}
\usepackage{booktabs}       
\usepackage{amssymb}
\usepackage{amsfonts}       
\usepackage{nicefrac}       
\usepackage{microtype}      
\usepackage{hyperref}

\usepackage{newfloat}
\usepackage{listings}
\DeclareCaptionStyle{ruled}{labelfont=normalfont,labelsep=colon,strut=off} 
\floatstyle{ruled}
\newfloat{listing}{tb}{lst}{}
\floatname{listing}{Listing}
\input{macros}

\title{Analysis and Predictive Modeling of Solar Coronal Holes Using Computer Vision and ARIMA-LSTM Networks}

\def\authorEmail{juyyun@cs.stonybrook.edu \\ This paper is accepted to the first joint European Space Agency SPAICE Conference 2024. \url{https://spaice.esa.int/}}

\author{
    Juyoung Yun\textsuperscript{\rm 1}\thanks{Corresponding author. E-Mail: \authorEmail}, 
    Jungmin Shin\textsuperscript{\rm 2}
}
\affiliations{
    \textsuperscript{\rm 1}Stony Brook University, New York, USA\\
    \textsuperscript{\rm 2}Defense Acquisition Program Administration (DAPA), Gwacheon-si, Republic of Korea
}

\usepackage{bibentry}

\usepackage{framed}

\begin{document}

\maketitle

\begin{abstract}
\input{Contents/0-abstract}
\end{abstract}

\section{Introduction}
\input{Contents/1-intro}

\section{Related Works}
\input{Contents/2-related}

\section{Methodology}
\input{Contents/3-method}

\section{Results}
\input{Contents/4-results}

\section{Discussion}
This study enhances understanding of coronal holes and their impacts on space weather by developing a detection and prediction model. While effective, the model focuses mainly on mid-zone areas, neglecting variations in regions like the polar regions which might influence space weather. Additionally, it covers only one solar cycle, limiting its scope. Future research should delve deeper into the physical processes governing coronal hole formation, particularly the sun's magnetic dynamics.

\section{Conclusion}
\input{Contents/5-conclusion}




\bibliography{aaai24}

\end{document}

%% file: macros.tex
\usepackage{microtype}
\usepackage{graphicx}
\usepackage{subfigure}
\usepackage{booktabs} 
\usepackage[linewidth=1pt]{mdframed}

\usepackage{amsmath}
\usepackage{amssymb}
\usepackage{mathtools}
\usepackage{amsthm}

\usepackage[capitalize,noabbrev]{cleveref}

\theoremstyle{plain}

\theoremstyle{definition}

\theoremstyle{remark}

\usepackage[textsize=tiny]{todonotes}


%% file: Contents/0-abstract.tex
In the era of space exploration, coronal holes on the sun play a significant role due to their impact on satellites and aircraft through their open magnetic fields and increased solar wind emissions. This study employs computer vision techniques to detect coronal hole regions and estimate their sizes using imagery from the Solar Dynamics Observatory (SDO). Additionally, we utilize hybrid time series prediction model, specifically combination of Long Short-Term Memory (LSTM) networks and ARIMA, to analyze trends in the area of coronal holes and predict their areas across various solar regions over a span of seven days. By examining time series data, we aim to identify patterns in coronal hole behavior and understand their potential effects on space weather.

%% file: Contents/1-intro.tex
The exploration of space has underscored the importance of understanding space weather phenomena, especially those related to coronal holes \cite{Davis, Sandford}. These regions on the sun, characterized by open magnetic field lines and cooler temperatures, emit solar winds at higher rates than their surroundings. Such emissions have profound implications for Earth, influencing operations of aircraft, satellites, and terrestrial technologies \cite{Filjar1, Rama}.

Space weather affects the expanse between the Sun and Earth, impacting the solar wind, Earth's magnetosphere, ionosphere, and thermosphere \cite{noaan}. Phenomena like solar flares and coronal mass ejections shape space weather, causing disruptions in satellite communications and GPS accuracy \cite{Schwenn, Rama}. While coronal holes' influence on space weather is well-known, predicting their size and evolution remains a challenge. Accurate prediction of coronal hole areas is crucial, as their size and position can directly affect the intensity and impact of solar wind streams \cite{Kappenman, Boteler}.

\begin{figure}
\centering
\includegraphics[width=1.0\columnwidth]{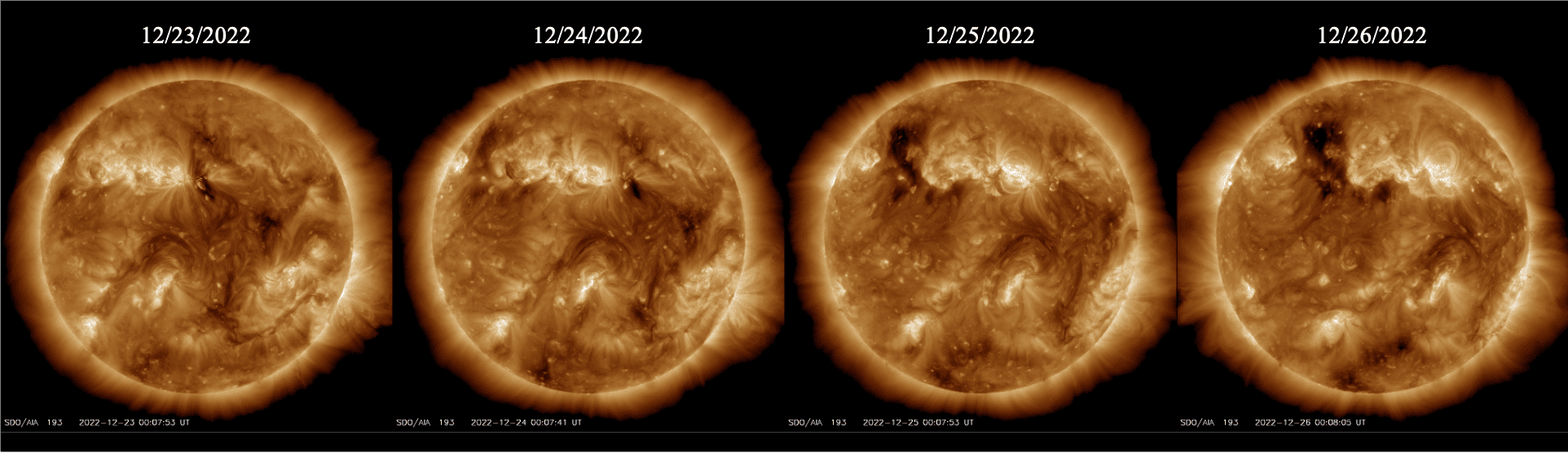}
\caption{\small A visual representation of the changes in coronal hole size and position over a 4-day period from December 23, 2022 to December 26, 2022~\cite{Nasa}. The image illustrates the dynamic nature of coronal holes and their potential impact on space weather events.}
\label{fig:aia}
\end{figure}

Addressing this gap, our study combines computer vision with deep learning to predict coronal hole sizes. Automated detection ensures accuracy and consistency, eliminating manual inspection errors, and enhancing efficiency. Using the Long Short-Term Memory (LSTM) model~\cite{Hochreiter} and ARIAM~\cite{arima}, we forecast coronal hole sizes over a week, providing insights into impending space weather events. Our scalable method allows for extensive dataset analysis, offering insights into long-term coronal hole patterns. This research highlights the critical role of AI in enhancing predictive capabilities for space weather events, ultimately protecting Earth's technological infrastructure \cite{noaan}.

%% file: Contents/2-related.tex
\begin{figure*}
\centering
\includegraphics[width=2.0\columnwidth]{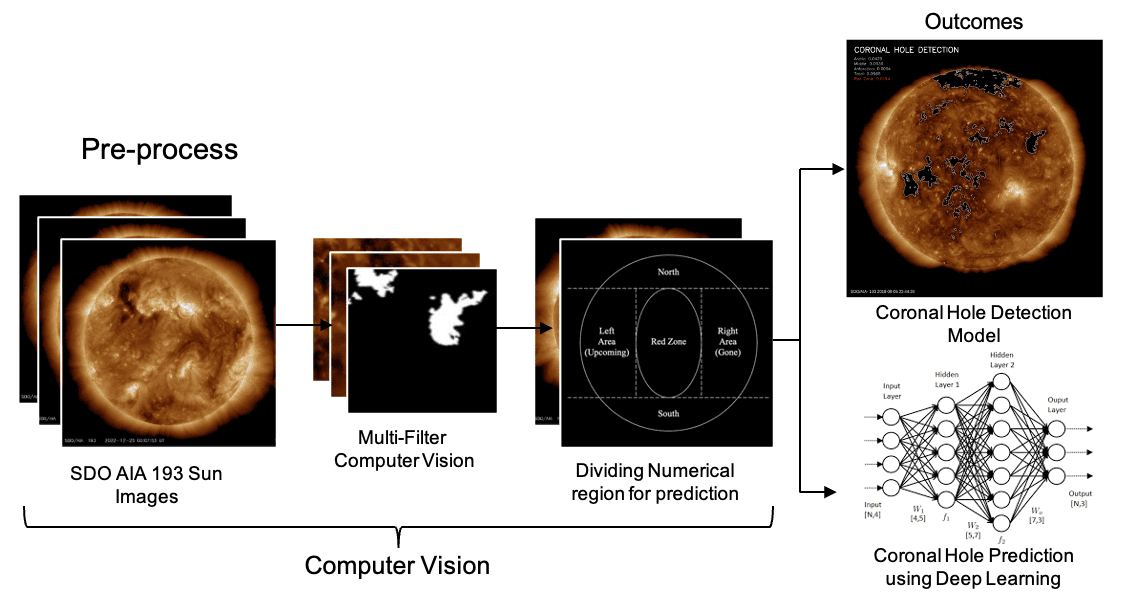}
\caption{Overall structure of the predictive modeling of coronal hole areas using computer vision and deep learning. The regions used to determine the coronal hole area on the sun are defined by the binary-based Coronal Hole Detection (BCH) model with AIA-193~\cite{Nasa}.}
\label{fig:sys}
\end{figure*}

In space weather studies, observing coronal holes is essential. These features are visible using extreme ultraviolet (EUV) wavelengths and are monitored by instruments like the AIA on the Solar Dynamics Observatory (SDO) \cite{Lemen, Nasa}. Computational models such as MHD and potential field source surface (PFSS) models help detect these holes by simulating the sun's magnetic field and plasma behavior \cite{Inoue}. Coronal mass ejections (CMEs) and corotating interaction regions (CIRs) from varying speed solar wind streams can induce geomagnetic storms, impacting Earth's space environment \cite{cme1, Schwenn, Heber, Mursula, Bobrov, Richardson1, Richardson2}. Geomagnetic activity is measured using indices like Kp, AE, and Dst, which help assess the intensity and effects of these storms \cite{Eoin, Filjar1, Rama, Badruddin, Bartels1, Bartels0, Cranmer, Noaa}. Historically, research has focused on detecting coronal holes using various techniques from manual inspection to advanced computer vision \cite{Linker2021, Jarolim2021}. While these methods improved detection accuracy, they did not predict future behavior. Our study addresses this gap by combining detection with prediction, using the LSTM~\cite{Hochreiter} model to forecast coronal hole area over a seven-day period. This advancement enhances our ability to prepare for space weather events in the future.

%% file: Contents/3-method.tex
We develop a novel approach to predict the area of coronal holes by utilizing computer vision and deep learning. Our methodology consists of two primary phases: detecting coronal holes through computer vision and predicting their future area using deep learning as depicted in Figure \ref{fig:sys}.

\begin{figure}[hbt!]
   \includegraphics[width=1.0\columnwidth]{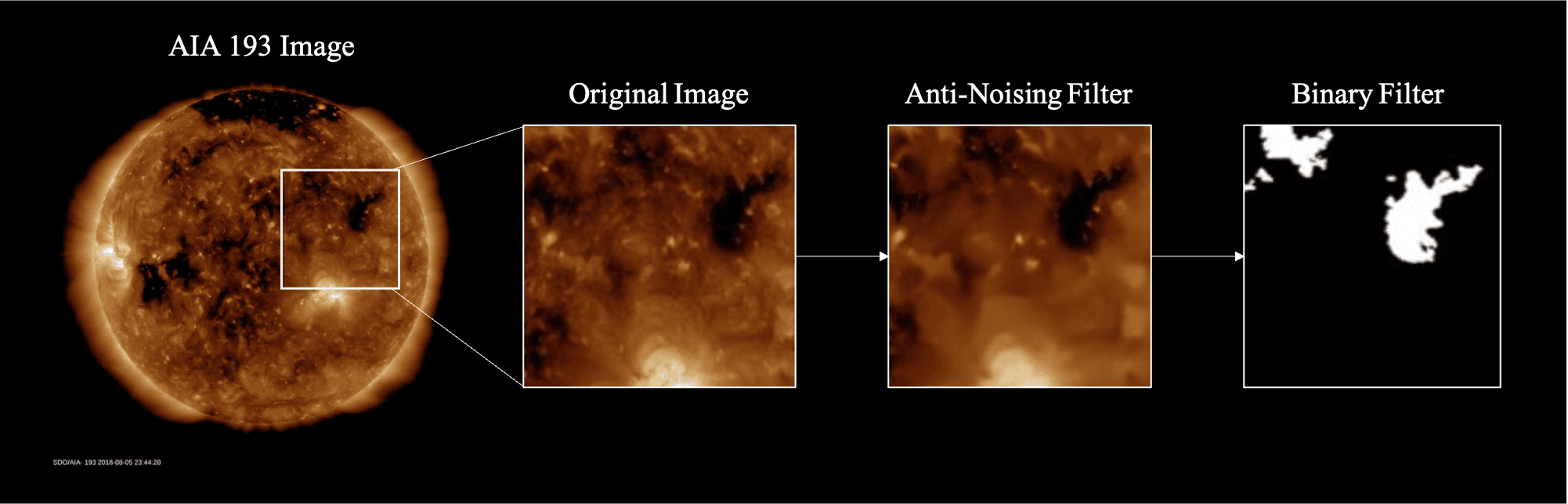}
   \caption{A depiction of the multi-filters computer vision based detection applied to an original AIA 193 image~\cite{Nasa}, including anti-noising and binary filters to accurately identify and measure coronal holes in the sun.} 
   \label{fig:filter}
\end{figure}

\subsection{Coronal Hole Detection using Computer Vision}


Our study employs advanced and simple computer vision techniques to automate the detection of coronal holes without any training deep learning model, utilizing methods like the non-local means (NLmeans) for enhancing image clarity and reducing noise, making features of coronal holes more distinguishable \cite{Buades} as shown in Figure~\ref{fig:filter}. Subsequently, we apply image binarization to transform the processed images into a binary format that highlights the coronal holes against the solar background \cite{Gonzalez}. This step is crucial for accurately delineating the boundaries of coronal holes. Finally, we refine our detection using the Laplacian of Gaussian (LoG) method, which is sensitive to abrupt changes in image intensity, aiding in precise contour detection of the coronal holes \cite{marr}. This comprehensive approach ensures accurate identification and analysis of coronal hole regions.

In our segmentation, the middle region, which includes the Left Area, Redzone, and Right Area, generally involves areas close to the solar equator, potentially extending to about ±30° latitude from the solar equator as shown in Figure~\ref{fig:sys}. These regions are more likely to impact Earth's geomagnetic environment directly due to their equatorial proximity.

Specifically, the Redzone is defined within a 15° wide longitudinal slice around the heliocentric meridian and falls within ±30° latitude from the solar equator~\cite{Nakagawa2019}. Areas outside of this middle region are classified as polar regions. Additionally, for quantitative analysis, we calculate the percentage of the white area (coronal holes) relative to the total solar surface area, considering the entire solar area as a unity 1.

\subsubsection{Coronal Hole Analysis}

\input{Contents/3-analysis}

\begin{figure*}
\centering
\includegraphics[width=2.0\columnwidth]{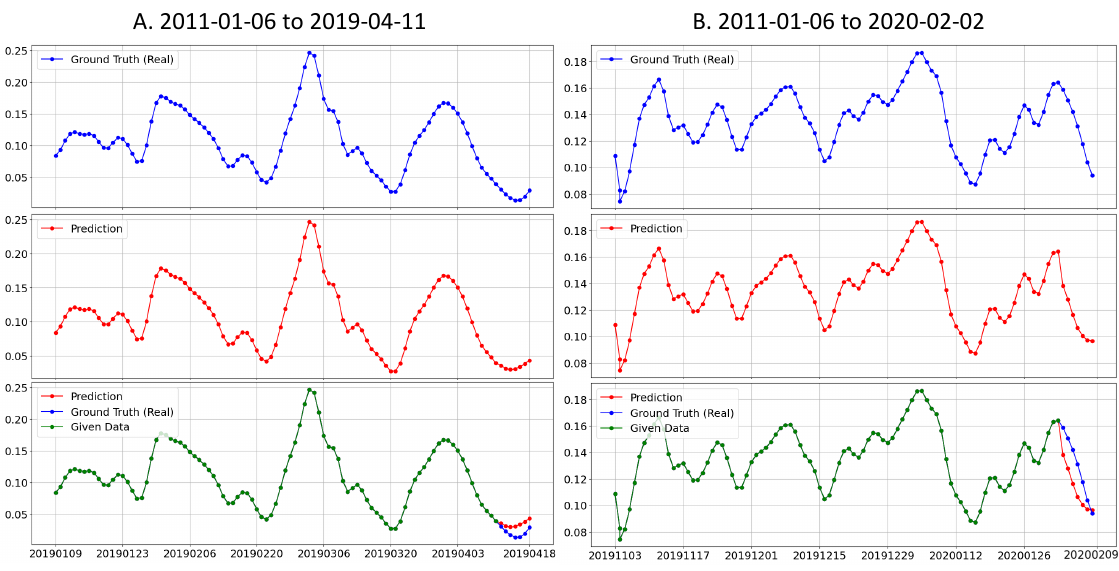}
\caption{The plots from A to B show the ARIMA-LSTM predicted the next 7 days of middle area of coronal holes. Each plot corresponds to a different time period used for training:
    (A) 2011-01-06 to 2019-04-11, 
    (B) 2011-01-06 to 2020-02-02.
    The blue lines indicate the actual coronal hole areas used for training and the actual areas for the following 7 days. The red lines show the predicted areas for the next 7 days based on the given training data. The green lines represent the data used for training. This visualization highlights the effectiveness of the ARIMA-LSTM hybrid model in predicting the future behavior of coronal holes.}
\label{fig:solar1}
\end{figure*}

\subsection{Time Series Prediction using Deep Learning}

In our study, we integrate Long Short-Term Memory (LSTM) networks~\cite{Hochreiter} with AutoRegressive Integrated Moving Average (ARIMA)~\cite{arima} models to predict the future behavior of coronal holes. LSTM networks efficiently manage long-term dependencies in data, crucial for accurate time series predictions~\cite{Yang, Torres}. Additionally, we use the pm.auto\_arima function from the pmdarima library~\cite{pmdarima} to automatically determine the best fitting parameters for the ARIMA model. For our predictions, we generate forecasts from both the ARIMA and LSTM models over a 7-day period and then compute the average of these predictions to obtain the final forecasted values. This hybrid approach enhances the overall prediction accuracy by leveraging the strengths of both models.

%% file: Contents/3-analysis.tex
Before diving into deep learning predictions, it's essential to focus on data extraction and preprocessing. Our work with the Binary-based Coronal Hole Detection (BCH) model ensures the data is accurate and meaningful, providing a strong foundation for deep learning applications.

\begin{figure}[hbt!]
   \centering
   \includegraphics[width=1.0\columnwidth]{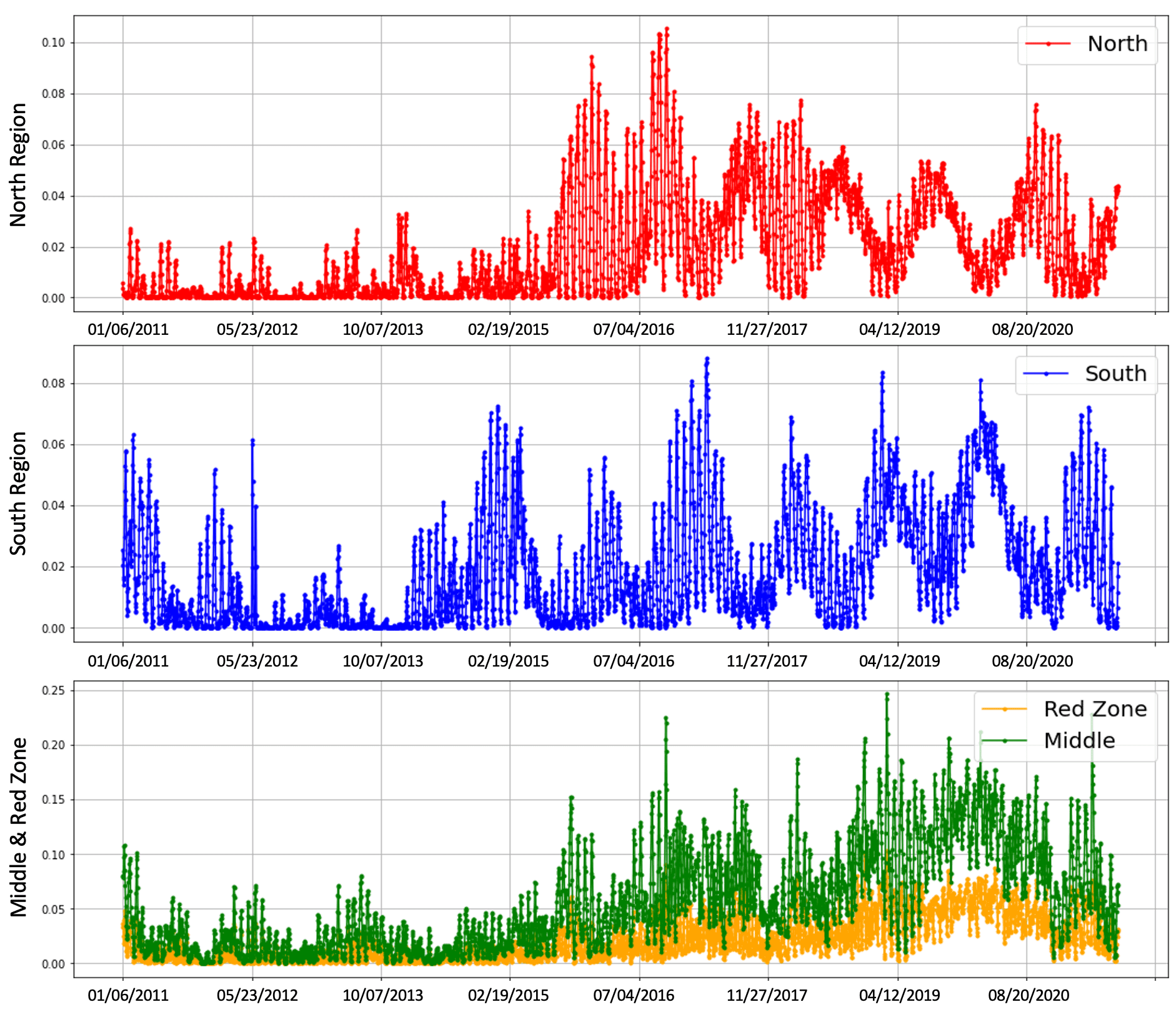}
   \caption{Daily coronal hole data for each region of the sun over 3857 days from January 6, 2011, to August 10, 2021, as detected by the BCH model. The graph shows the variation in coronal hole area over time.} 
   \label{fig:regions}
\end{figure}

\textbf{Overall.} Figure \ref{fig:regions} shows the expansion of coronal hole areas over time, particularly from January 6, 2011, to August 10, 2021. The middle region of the sun has seen a significant increase, suggesting a trend that requires longer-term data for a comprehensive understanding.

\begin{figure}[hbt!]
    \centering
   \includegraphics[width=1.0\columnwidth]{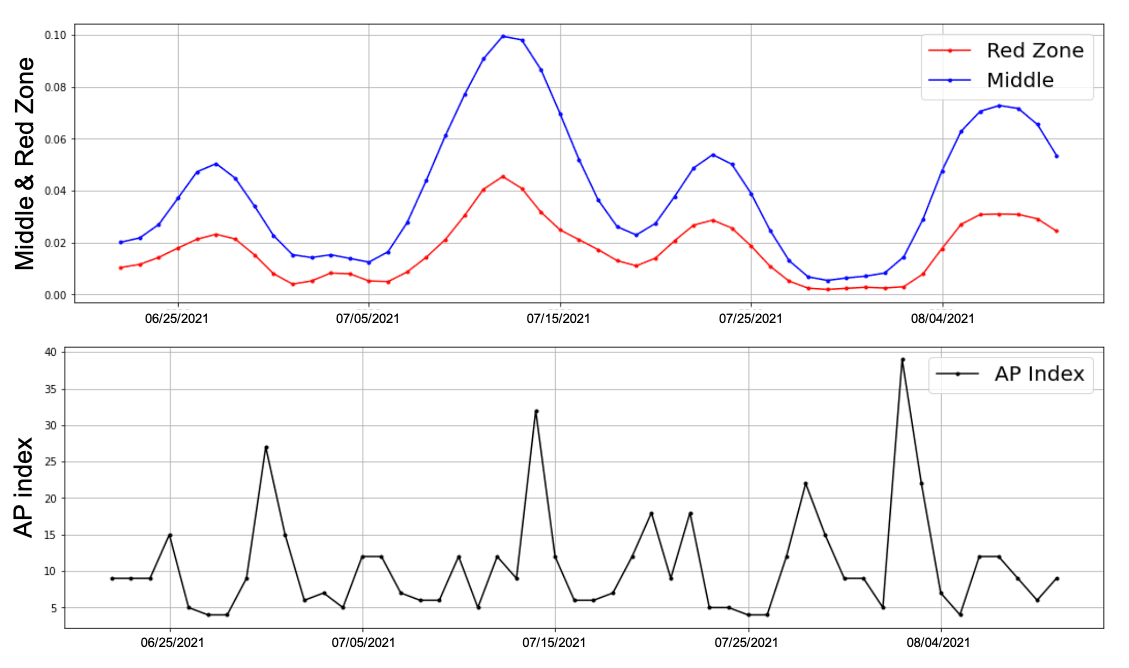}
   \caption{The AP index and coronal hole areas in the middle and redzone regions of the sun from June 22, 2021, to August 10, 2021, as detected by the BCH model. The graph shows the correlation between the AP index and coronal hole area in these regions.} 
   \label{fig:midred}
\end{figure}

\textbf{Middle Region.} Our analysis indicates a correlation between the size of coronal holes in the middle region and redzone and geomagnetic disturbances on Earth, as measured by the AP index. Figure \ref{fig:midred} highlights this relationship over a 50-day period, showing that increases in the size of the middle region often precede rises in the AP index. This finding is consistent with previous research suggesting that severe coronal holes in the middle region can impact Earth's geomagnetic environment within 3-5 days~\cite{Watari2023}.

\begin{figure}[hbt!]
    \centering
    \includegraphics[width=1.0\columnwidth]{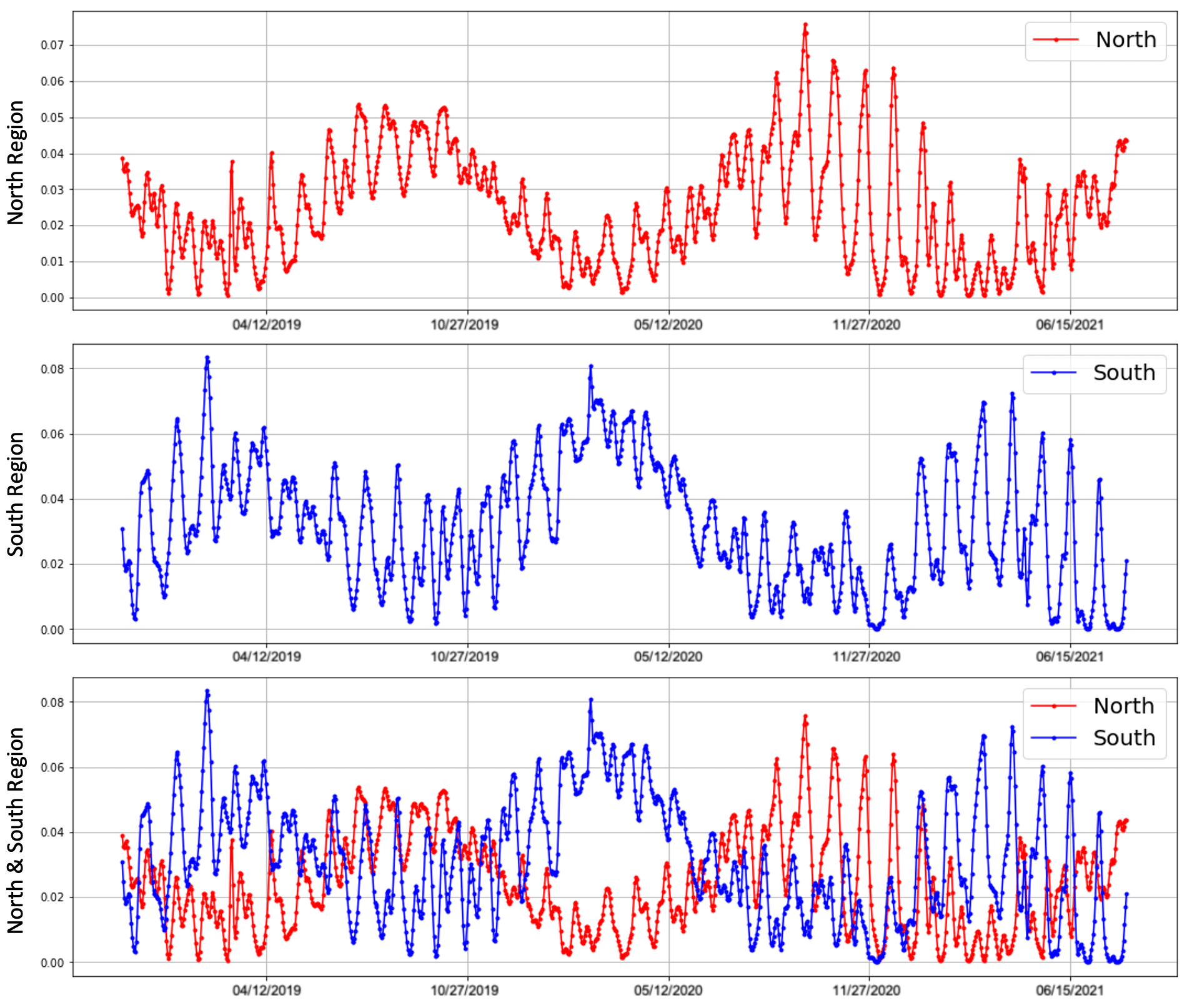}
   \caption{Daily coronal hole data for the sun's south and north regions over 1000 days from November 13, 2018, to August 10, 2021, as detected by the BCH model. The graph shows the variation in coronal hole area and the consistency of the polarity rule.}
   \label{fig:poles}
\end{figure}

\textbf{Polar Region.} The polar regions of the sun also exhibit distinct behaviors, as shown in Figure \ref{fig:poles}. The "polarity rule" demonstrates that coronal hole areas at the north and south poles often behave oppositely, aligning with their magnetic field configurations and affecting solar wind distribution. These findings emphasize the importance of continuous monitoring and further research to understand the long-term impacts of coronal holes on space weather and geomagnetic disturbances on Earth

%% file: Contents/4-results.tex
\subsection{Experimental Setting}
In our study, we employed a Long Short-Term Memory (LSTM) network to predict the future behavior of coronal holes. The LSTM model architecture consists of an initial LSTM layer with 50 units and a ReLU activation function, configured to return sequences. This is followed by a Dropout layer with a dropout rate of 0.2 to prevent overfitting. Next, there is another LSTM layer with 50 units and a ReLU activation function, which does not return sequences. Another Dropout layer with a dropout rate of 0.2 is added, and the final layer is a Dense layer with a single unit for the output. The model is compiled using the mean squared error (MSE) loss function and the Adam optimizer, with the learning rate set to the default value. The model was trained for 50 epochs with a batch size of 128 and a learning rate of 0.01.

\subsection{Experimental Results}
We employ ARIMA-LSTM~\cite{Hochreiter, arima} networks to predict the area of coronal holes in the sun's middle region. The LSTM model, trained on 3857 days of numerical data from the middle regions of AIA 193 observations extracted by the BCH model, is adept at handling time-series data, capturing patterns and trends in solar activity.

\begin{figure*}
\centering
\includegraphics[width=2.0\columnwidth]{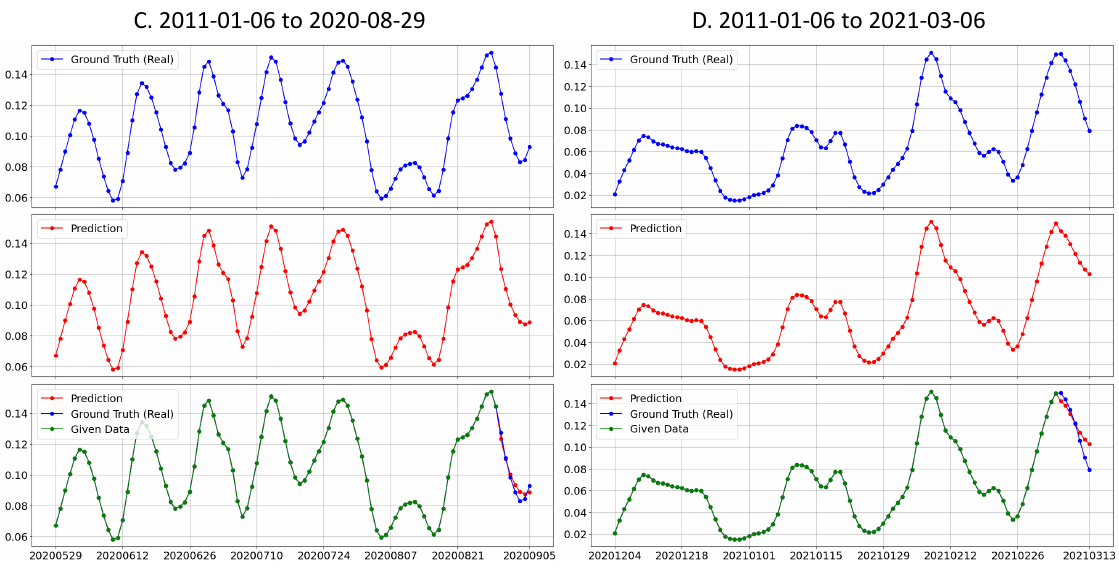}
\caption{The plots from C to D show the ARIMA-LSTM predicted the next 7 days of middle area of coronal holes. Each plot corresponds to a different time period used for training:
    (C) 2011-01-06 to 2020-08-29, 
    (D) 2011-01-06 to 2021-03-06.
    The blue lines indicate the actual coronal hole areas used for training and the actual areas for the following 7 days. The red lines show the predicted areas for the next 7 days based on the given training data. The green lines represent the data used for training. This visualization highlights the effectiveness of the ARIMA-LSTM hybrid model in predicting the future behavior of coronal holes.}
\label{fig:solar2}
\end{figure*}

Figure \ref{fig:solar1}, \ref{fig:solar2} illustrates the performance of the ARIMA-LSTM hybrid model in predicting the Daily middle area of coronal holes for the next 7 days. The model is trained on different time periods, as shown in plots A to D, which cover various spans from 2011-01-06 to 2021-03-06. The blue lines represent the actual coronal hole areas used for training and the actual areas for the subsequent 7 days. The red lines indicate the predicted areas for the next 7 days based on the training data, and the green lines represent the data used for training.

The patterns observed in the coronal hole areas suggest that these features exhibit specific trends and behaviors over time. By leveraging these patterns, the ARIMA-LSTM hybrid model can effectively learn and predict future coronal hole areas. This approach demonstrates that it is possible to accurately predict the behavior of coronal holes based on their historical patterns, providing valuable insights for understanding and forecasting space weather phenomena.

%% file: Contents/5-conclusion.tex
A simplified model for detecting coronal holes is essential for enhancing space weather forecasting. Coronal holes, marked by their low temperature and density, influence solar winds that can affect Earth's satellites and power grids. Accurate detection and measurement of coronal holes are crucial for predicting space weather events and mitigating their impacts. Our research developed a model that identifies coronal holes, generating 3857 data entries. This data was used to create a deep learning prediction model, specifically using ARIMA-LSTM, to estimate the size of coronal holes. The model showed promising results, effectively predicting the coronal hole size using numerical data. This model improves our ability to predict space weather events, aiding in the prevention of potential impacts on Earth's technological infrastructure. Additionally, understanding the size and location of coronal holes helps in predicting solar activities like CMEs and CIRs.